\documentstyle[11pt]{article}
\title{\begin{flushright}
{\small NSF-ITP-93-127 \\
NUC-MINN-93/25-T \\
SUNY-NTG-93-45 \\
October 1993 \\}
\end{flushright}
{\bf WEINBERG-TYPE SUM RULES AT \\ ZERO AND FINITE TEMPERATURE}}
\author{{\bf J. I. Kapusta}$^{\dagger}$ and
 {\bf E. V. Shuryak}$^{\ddagger}$ \\
  {\it Institute for Theoretical Physics} \\
  {\it University of California} \\
  {\it Santa Barbara, CA 93106}}

\date{}

\begin{document}

\maketitle

\begin{center}
Abstract\\
\end{center}

We consider sum rules of the Weinberg type at zero and nonzero temperatures.
On the basis of the operator product expansion at zero temperature we
obtain a new sum rule which involves the average of a four-quark operator
on one side and experimentally measured spectral densities on the other.
We further generalize the sum rules to finite temperature.  These
involve transverse and longitudinal spectral densities at each value of the
momentum.  Various scenarios for the relation between chiral symmetry
restoration and these finite temperature sum rules are discussed.

\vspace{.1in}

\noindent $^{\dagger}$ On leave from: {\it School of Physics and Astronomy,
University of Minnesota, Minneapolis, MN 55455}\\
\noindent $^{\ddagger}$ On leave from: {\it Physics Department,
State University of New York, Stony Brook, NY 11794}

\vfill \eject

\section{Introduction}

In a famous 1967 paper \cite{Weinberg} Steven Weinberg asked the question:
``What relations are imposed by current algebra upon the spectra of
the 1$^+$ and 1$^-$ mesons?"  Under certain conditions the answer was
two sum rules involving the vector and axial-vector spectral densities.
They are known as the Weinberg sum rules:
\begin{eqnarray}
& {\rm I} & \;\;\;\;\;\;\;\;\;\;\;\;\;\;\;\;\;
\int_0^{\infty}\frac{ds}{s} \, [\rho_V(s)-\rho_A(s)]\,=\,F^2_{\pi} \,
 , \;\;\;\;\;\;\;\;\;\;\;\;\;\;\;\;\;\;\;\;\;\;\;\;\;\;\;\;\;\; \\
& {\rm II} & \;\;\;\;\;\;\;\;\;\;\;\;\;\;\;\;\;\;
\int_0^{\infty}ds \, [\rho_V(s)-\rho_A(s)]\,=\,0 \, .
\end{eqnarray}
Assuming vector meson dominance and the KSFR relation these sum rules
lead to the prediction that the $\rho$ and $a_1$ masses are related by
$m_{a_1} = \sqrt{2}m_{\rho}$, which is approximately valid.  In this paper
we ask two questions.  The first one is:  Given that QCD is now known to
be the theory of the strong interactions, what extra information can
we get from sum rules of the Weinberg type?

The last fifteen years has
seen a great deal of activity surrounding QCD at finite temperature.
Of particular interest are the issues of deconfinement and chiral
symmetry restoration at temperatures of the order of 160 MeV.
Therefore, we are led to consider a second question:  What are the
implications of the approach to chiral symmetry restoration at finite
temperature for sum rules of the Weinberg type?

The status of the original Weinberg sum rules in the context of QCD
sum rules was discussed by Shifman, Vainshtein and Zakharov \cite{SVZ}
and then by Narison \cite{Narison}, while a more up-to-date
phenomenological analysis was performed by Peccei and Sola \cite
{Peccei_Sola}.  The two sum rules derived by Weinberg are very general,
as he showed, and do not depend on specific details of the QCD
Lagrangian.  Higher order sum rules (involving more powers of $s$ in
the integrand) do depend on dynamics of chiral symmetry breaking
in the vacuum.  In section 2 we derive a third sum rule of the
type of eqs. (1-2).  This new sum rule involves the vacuum expectation
value (VEV) of a certain local four-quark operator.  It can be
obtained from the sum rule if we know the vector and axial-vector
spectral densities accurately enough from experiment.  It can
also be obtained from lattice QCD; the chirality-violating structure
of the operator helps here because its VEV has no short distance
perturbative contribution.  We perform a detailed analysis of all
sum rules in section 3.  We will see that they are restrictive enough
to fill in gaps in the experimental data, allowing us to determine
the spectral densities with quite some accuracy.

There has been a lot of discussion in the literature and at conferences
about the temperature dependence of hadron masses.  Some calculations
yield increasing masses, some yield decreasing masses, and still others
yield masses that either increase or decrease depending on the quantum
numbers of the hadron; see
\cite{Bochkarev,Brown_Rho,Shuryak_pot,Hatsuda,Song} and the review
\cite{Shuryak_cor}.
Clearly all these calculations are only approximate.
In addition, the very notion of a mass at finite temperature must be
very clearly defined, such as the screening mass or the pole mass
corresponding to collective excitations.

A common denominator of all studies of this type is the temperature
dependence of correlation functions.  It would be good if some general
statements about these correlation functions could be made which rely
on the fundamental properties of QCD at finite temperature.  This is
the aim in section 4.  We generalize the original Weinberg sum rules,
and the new one,
to finite temperature.  The first one (eq. (1)) generalizes to a sum
rule involving only the longitudinal spectral density and depends on
three-momentum.  The second one (eq. (2)) generalizes to two separate
sum rules, one involving the longitudinal spectral density and the other
involving the transverse spectral density, both depending on momentum.
At zero three-momentum they collapse to the same expression.  In the
vacuum there is no dependence on momentum because of Lorentz invariance,
but at finite temperature there is a preferred rest frame, hence a
dependence on momentum and on polarization.  We would like to point out
here that probably the first discussion of Weinberg sum rules at finite
temperature was given by Bochkarev and Shaposhnikov \cite{Bochkarev}
in the context of QCD sum rules and for zero momentum.

The finite temperature sum rules can be used to constrain models or
approximations to QCD, and can help us to understand the approach to
chiral symmetry restoration.  Various possibilities will be considered
in section 5.  We should reference here the early paper on phenomenology
of the chiral phase transition in heavy ion collisions by Pisarski
\cite{Rob}. For recent discussion of the topic one can see \cite{Shuryak_U(1)}.

We remark that throughout this paper we assume that the up and down
quark masses are identically zero so that chiral symmetry is exact.
Consideration of the impact of nonzero quark masses on the original
Weinberg sum rules within perturbative QCD was done by Floratos,
Narison and de Rafael \cite{FNR}.

\section{Derivation of Zero Temperature Sum Rules from QCD}

We define the vector and axial-vector currents,
\begin{eqnarray}
V_\mu^a=\bar q \gamma_\mu (\tau^a/2) q \, , \\
A_\mu^a=\bar q \gamma_\mu \gamma_5 (\tau^a/2) q \, ,
\end{eqnarray}
where $\tau^a/2$ is the isospin generator.  With this normalization
the current algebra of charges obeys the equal time commutation relations
\begin{eqnarray}
\left[Q^a_V,Q^b_V\right] & = & i\varepsilon^{abc}Q^c_V \, ,\\
\left[Q^a_V,Q^b_A\right] & = & i\varepsilon^{abc}Q^c_A \, ,\\
\left[Q^a_A,Q^b_A\right] & = & i\varepsilon^{abc}Q^c_V \, .
\end{eqnarray}
We define the vector and axial-vector spectral densities
in the usual way.  They are positive definite quantities
defined for positive $s$.
\begin{equation}
<0|V_a^{\mu}(x) V_b^{\nu}(0)|0> \, = \, -\frac{\delta^{ab}}{(2\pi)^3}
\int d^4p \, \theta(p^0) \, e^{ip\cdot x}\left(g^{\mu\nu}-\frac{p^{\mu}
p^{\nu}}{p^2} \right) \rho_V(s) \, ,
\end{equation}
\begin{eqnarray}
\lefteqn{<0|A_a^{\mu}(x) A_b^{\nu}(0)|0> \, = \nonumber} \\
&&-\frac{\delta^{ab}}{(2\pi)^3}
\int d^4p \, \theta(p^0) \, e^{ip\cdot x}
\left[ \left(g^{\mu\nu}- \frac{p^{\mu}p^{\nu}}{p^2} \right) \rho_A(s) +
F_{\pi}^2 \delta(s) p^{\mu}p^{\nu} \right] \, .
\end{eqnarray}
The dimension of the spectral densities is energy-squared.  Note the
explicit contribution of the pion to the axial-vector correlator.

In this paper we work in imaginary time so that all distances are
space-like, or Euclidean: $x^2=t^2-r^2=-\tau^2$.  In this domain
the spectral representation of the correlation functions looks as follows
\cite{Shuryak_cor}:
\begin{displaymath}
\Delta D^{ab\mu}_{\mu}(\tau) \, \equiv \,
<0|{\cal T} \left[ V^{a\mu}(x) V^b_{\mu}(0) \, - \,
A^{a\mu}(x) A^b_{\mu}(0) \right] |0> \, =
\end{displaymath}
\begin{equation}
- \frac{\delta^{ab}}{4\pi^2\tau} \int_0^{\infty} ds \, \sqrt{s} \,
\left[ 3\rho_V(s) - 3\rho_A(s) - s \, F_{\pi}^2 \delta(s) \right]
K_1(\sqrt{s}\tau) \, ,
\end{equation}
and
\begin{displaymath}
\Delta D^{00}_{ab}(\tau) \, \equiv \,
<0|{\cal T} \left[ V_a^0(x) V_b^0(0) \, - \,
A_a^0(x) A_b^0(0) \right] |0>  \, = \,
- \frac{\delta_{ab}}{4\pi^2\tau} \int_0^{\infty} ds \, \sqrt{s} \,
\end{displaymath}
\begin{equation}
\left[ \rho_V(s) - \rho_A(s) - s \, F_{\pi}^2 \delta(s) \right]
\left[ \frac{K_0(\sqrt{s}\tau)}{\sqrt{s}\tau} + \left(\frac{2}{s\tau^2}
+ 1 \right)K_1(\sqrt{s}\tau) \right] \, .
\end{equation}
Notice that the integrands essentially involve the
standard Feynman propagator for a particle of mass $m$
which, in the Euclidean domain, is
\begin{eqnarray}
D(m,\tau)_{\rm free \; scalar}\,=\,{m \over 4\pi^2 \tau} K_1(m\tau) \, .
\end{eqnarray}
Exponential decay of the Bessel function $K_1$ at large argument ensures
convergence of such integrals for any QCD correlation functions,
except probably at $\tau=0$. In this sense, there is no difference between
the Euclidean time representation
\cite{Shuryak_cor} and the Borel-transformed sum rules \cite{SVZ}, in which
the propagator is replaced with $\exp(-s/M^2)$, with Borel parameter
$M$ replacing Euclidean time $\tau$.

The coordinate representation is more transparent and accessible to
numerical methods, such as lattice calculations.  Recent studies based
on the instanton liquid model \cite{Shuryak_Verbaarschot_cor}
and lattice QCD \cite{Negele_etal} have reported on the
calculation of a set of Euclidean correlation functions, including vector and
axial ones. Unfortunately, none of them has focused on their {\it difference}
with sufficiently high accuracy, and therefore they are not discussed in the
present work.

We now come to the central idea behind the derivation of the sum rules:
{\it each} sum rule corresponds to
a {\it particular term} in the small-distance
asymptotic expansion of the correlation function.

In the limit $\tau \rightarrow 0$ the
product of currents can be expanded according to the
operator product expansion (OPE), a very powerful means for
connecting VEV's of quark and gluon operators to experimentally
observable hadronic properties.  The first
terms in this expansion were first computed in \cite{SVZ}.
For the contracted polarization tensor the result is
\begin{displaymath}
D^{ab\mu}_{\mu}(\tau) \, \equiv \,
<0|{\cal T} \left[ V^{a\mu}(x) V^b_{\mu}(0) \right] |0> \, =
\end{displaymath}
\begin{equation}
-\frac{3\delta^{ab}}{\pi^4\tau^6}\left[ 1 + \frac{\alpha_s(\tau)}
{\pi} - \frac{<0|\left(gF_{\mu\nu}^c\right)^2|0>\tau^4}{3\cdot 2^7}
-\frac{\pi^2\tau^6}{8} \ln(\mu\tau) <0|{\cal O}_{\rho}|0> + \cdots \right] \,
\end{equation}
where, in the argument of the logarithm, $\mu << 1/\tau$ is the
renormalization scale, and ${\cal O}_{\rho}$ is a complicated
four-quark operator.  There
is a similar expression for the correlator of two axial-vector currents
but with a different four-quark operator ${\cal O}_{a_1}$.  For our
purposes we only need their difference, which is given below.

Since chiral symmetry breaking is a long wavelength phenomenon,
at very short distances, or at very high energies, the difference
between vector and axial-vector correlators should go to zero.
Indeed, taking this difference one finds that all terms except
for the four-quark operators in eq. (13) drop out.

One can now look for consequences of this statement for the spectral density.
Expanding the Bessel function in eq. (10) for small values of $\tau$ we get
\begin{eqnarray}
\Delta D^{ab\mu}_{\mu}(\tau) & = &
- \frac{3\delta^{ab}}{4\pi^2} \int_0^{\infty} ds
\left[ \rho_V(s) - \rho_A(s) \right] \nonumber \\
&& \left[ \frac{1}{\tau^2} + \frac{s}{2}\ln \left( \frac{\sqrt{s}\tau}
{2} e^{C-1/2} \right) + {\rm order}\left(\tau^2,\tau^2\ln\tau \right)
\right] \, ,
\end{eqnarray}
where $C$ is Euler's constant.
The OPE has no power divergence in $\tau$ in the {\it difference}
$\Delta D^{ab\mu}_{\mu}$.  Therefore the coefficient of $1/\tau^2$
in eq. (14) must vanish.  This is just the second Weinberg sum rule
(eq. (2)).  In the OPE framework it simply follows from the observation
that the first covariant operators which are not chirality blind
are four-quark ones which have dimension 6 or more.  Similarly
expanding eq. (11) for small $\tau$ and applying the observation
of chirality blindness we get
\begin{equation}
\int_0^{\infty} \frac{ds}{s}
\left[ \rho_V(s) - \rho_A(s) - s \, F_{\pi}^2 \delta(s) \right]
\left[ \frac{1}{\tau^4} + \frac{s}{4\tau^2} \right] \, = \, 0.
\end{equation}
The first and second terms in the last square brackets reproduce
the first and second Weinberg sum rules, respectively.

The next term in the small $\tau$ expansion is logarithmic.  In eq. (14)
we multiply the argument of the logarithm by $\mu/\mu$ which we must
do to match the OPE.  Equating the coefficients of $\ln(\mu\tau)$
in $\Delta D^{ab\mu}_{\mu}(\tau)$ we obtain the third sum rule,
\begin{equation}
{\rm III} \;\;\;\;\;\;\;\;\;\;\;
\int_0^{\infty} ds\,s \, \left[ \rho_V(s) - \rho_A(s) \right]
\, = \, - 2\pi <0|\alpha_s{\cal O}_\mu^\mu |0> \, .
\,\,\,\,\,\,\,\,
\end{equation}
Here
\begin{equation}
{\cal O}^{\mu\nu} \, = \,
\left( \bar u_L \gamma^\mu t^a u_L -\bar d_L \gamma^\mu t^a d_L \right)
\left( \bar u_R \gamma^\nu t^a u_R -\bar d_R \gamma^\nu t^a d_R \right) \, ,
\end{equation}
where $t^a$ are the color SU(3) matrices and R, L stand for right
and left-handed quarks.  Note the appearance of the renormalization
scale $\mu$ on the right side of this sum rule.  Since the other
side of the equation is expressed in terms of physical observables,
it must be that $\alpha_s(\mu)$ times the four-quark operator is a
renormalization group invariant.

The numerical value of the VEV of this operator is unknown. The
estimate suggested in \cite{SVZ} is based on
the so called ``vacuum dominance" hypothesis, which leads to
\begin{equation}
<0|{\cal O}_\mu^\mu |0> \, = \, {8 \over 9}
<0|\bar u u|0>^2 \, .
\end{equation}
The accuracy of this estimate should of course be
questioned, and various models of chiral symmetry breaking \cite{Shuryak_82}
and lattice numerical calculations can be used for that purpose. Let us
only add a comment on $\mu$-dependence here. If the vacuum dominance
hypothesis is correct, then the VEV should be proportional
to $[\ln(\mu/\Lambda_{QCD})]^{8/b}$,
the anomalous dimension of the quark condensate. (Here
b=${11 \over 3} N_c - {2\over 3} N_f
$ comes from the Gell-Mann-Low function.) Since the power is close to 1,
after being multiplied by $\alpha_s(\mu)\sim 1/\ln(\mu/\Lambda_{QCD})$
the right side of the third sum rule is nearly $\mu$-independent.
Thus, at least concerning the $\mu$-dependence,
this approximation can approximately hold.

The regular ($\tau$-independent) term was not considered in the QCD sum rule
context; it was first discussed in connection with point-to-point correlators
in the coordinate representation by one of us \cite{Shuryak_cor}.
It is interesting to express it in terms of an integral over the
difference in spectral densities, and it may be useful for lattice
calculations.  Dropping terms which vanish in the limit, we find
\begin{eqnarray}
\Delta D^{\mu}_{\mu}(\tau \rightarrow 0)&=&
- \ln(\mu\tau) \frac{3}{8\pi^2} \int_0^{\infty} ds\,s
\left[ \rho_V(s) - \rho_A(s) \right] \nonumber \\
&-& \frac{3}{8\pi^2} \int_0^{\infty} ds\,s
\, \ln\left(\frac{\sqrt{s}}{\tilde\mu} \right)
\left[ \rho_V(s) - \rho_A(s) \right] \, ,
\end{eqnarray}
where $\tilde\mu = 2\mu e^{1/2-C} = 1.85\mu$.  The use of $\mu$ here
is just for convenience; $\Delta D$ is actually independent of it.

\section{Phenomenology at Zero Temperature}

Phenomenological analysis of the Weinberg sum rules was originally made
in a very simple approximation using only the contributions of
$\rho, a_1, \pi$ mesons.  In other words, Weinberg {\it assumed} that
contributions from all excited states other than the lowest resonances
mentioned cancelled out.  Together with the KSFR relation it leads to the
famous prediction $m_{a_1}=\sqrt{2} m_\rho$ which looked excellent
from the point of view of data available at the time. However now we know
that this prediction, as well as predictions for coupling constants,
agrees with experiment only up to the level of 10-20\%.

The sum rules are exact in the chiral limit, so one should be willing
to verify them as accurately as possible.
If the complete spectral densities were measured one could simply evaluate
the integrals and check whether the sum rules are indeed satisfied,
up to the accuracy of the data.  Unfortunately the situation is not that
straightforward because there are {\it no} meaningful measurements of
the non-resonance contribution in the axial channel. Therefore
we first have to close this hole using the sum rules themselves.

Let us first discuss how well the spectral densities are determined
experimentally.  In the pole plus continuum approximation one would write
\begin{equation}
\rho_V(s) \, = \, \frac{m_{\rho}^4}{g_{\rho}^2} \delta(s-m^2_{\rho})
\, + \, \frac{s}{8\pi^2} \left[1\, + \, \frac{\alpha_s(s)}{\pi} \,
+ \cdots \right]\theta(s-E_V^2) \, ,
\end{equation}
and
\begin{equation}
\rho_A(s) \, = \, \frac{m_{a_1}^4}{g_a^2} \delta(s-m^2_{a_1})
\, + \, \frac{s}{8\pi^2} \left[1\, + \, \frac{\alpha_s(s)}{\pi} \,
+ \cdots \right]\theta(s-E_A^2) \, .
\end{equation}
The continuum is the same in the vector and axial-vector channels,
according to perturbative QCD, but the phenomenological threshhold is
in general different.  Note that the individual integrals over $s$ for
the vector and axial-vector channels which enter the sum rules are
actually divergent because of the continuum, but the {\it difference}
is finite.  The coupling constants are the same ones used in a
vector dominance approximation to the currents as expressed in the
current-field identities of Sakurai,
\begin{equation}
V_{\mu}^a \, = \, \frac{m_{\rho}^2}{g_{\rho}} \rho_{\mu}^a \, ,
\end{equation}
\begin{equation}
A_{\mu}^a \, = \, \frac{m_{a_1}^2}{g_a} a_{\mu}^a \, + \, {\rm pion} \, .
\end{equation}
We don't use these approximations for the spectral densities because
the three sum rules involve integrations of the spectral densities with
different powers of s and so it is likely
important to incorporate the finite widths of the resonances.

The vector spectral density is very well measured in $e^+e^- \rightarrow
\rho \rightarrow \pi^+\pi^-$.  An s-wave relativistic Breit-Wigner
is not a good representation because the $\rho$ meson is a p-wave
resonance.  A much better representation is given by the Gounaris-Sakurai
formula \cite{GS,GK1}.  It turns out that this complicated formula can
be approximated by a relativistic Breit-Wigner with an effective width
$\Gamma'_{\rho}$ = 118 MeV and an effective pole mass
$m'_{\rho}$ = 761 MeV.
\begin{eqnarray}
\rho_V(s) &=& {m^4_\rho\over g_\rho^2}{1\over\pi}
{m_\rho \Gamma'_{\rho} \over
(s-m'^2_\rho)^2+m^2_\rho \Gamma'^2_{\rho}} \nonumber \\
&+& {s\over 8\pi^2}\,{1 \over 1+ \exp[(E_V-\sqrt{s})/\delta_V]}
\left[1+ {0.22 \over \ln(1+\sqrt{s}/0.2 \, {\rm GeV})} \right] \, .
\end{eqnarray}
We take $g_{\rho}$ from the KSFR relation
\begin{equation}
g_{\rho}^2 \, = \, \frac{m_{\rho}^2}{2F_{\pi}^2} \, .
\end{equation}
With $m_{\rho}$ = 768 MeV and $F_{\pi}$ = 94.5 MeV one gets from this
$g_{\rho}^2/4\pi$ = 2.63.
The second term in eq. (24), corresponding to the continuum
from 2n-pion states (n = 2, 3, ...), has $E_V = 1.3$ GeV and
$\delta_V$ = 0.2 GeV \cite{Shuryak_cor}.

The coupling of the $a_1$ to the current can be determined from the
measured branching ratios of $\tau\rightarrow \nu_\tau+{\rm hadrons}$.
According to the 1992 Particle Data Table, the branching into the two
3-pion channels dominated by the $a_1$ is 11.2 $\pm$ 1.4\% while the
$\rho$ dominated 2-pion channel is 24.0 $\pm$ 0.6\%. The first number
gives rise to the main uncertainty in our numerical analysis below.
Using these numbers and the theoretical expression for the branching ratios
(which follows from the narrow width approximation)
\begin{equation}
{B(\tau \rightarrow \nu_\tau + a_1) \over
B(\tau \rightarrow \nu_\tau + \rho)} =
{m_{a_1}^2 \over m_{\rho}^2} {g^2_{\rho} \over g^2_{a_1}}
 {(1-m^2_{a_1}/m^2_\tau)^2 \over
 (1-m^2_\rho/m^2_\tau)^2} {(1+2m^2_{a_1}/m^2_\tau) \over
(1+2m^2_\rho/m^2_\tau)}
\end{equation}
one can get the coupling $g_{a_1}$ = 10.5 $\pm$ 0.7.

For the axial-vector spectral density we use an expression analogous
to the vector one but with the following differences.  First, we use
a constant width of 400 MeV and a constant mass of 1260 MeV for the
$a_1$ contribution (for more details about this see reference \cite{Isgur}).
We have, however, cut off this resonance below the threshold $m_\rho+m_\pi$.

The large width of the $a_1$ and its proximity to the $\tau$ lepton
causes a significant correction to eq. (26).  Numerically
integrating the differential decay rate \cite{Peccei_Sola} with the
realistic shape of the resonance we get finally a value
$g_{a_1}$ = 9.1 $\pm$ 0.7.

The available data for the nonresonant axial states are very poor
so that the continuum threshhold $E_A$ and the width $\delta_A$ are unknown.
The reason is partly statistical.  More importantly, since the data
about the axial spectral density come from the $\tau$ lepton decay,
there are fundamental limitations due to the $\tau$  mass which is not
big enough to provide sufficient phase space for 3- and 5-pion final states
with the needed invariant mass.
Therefore, some authors (for example \cite{Peccei_Sola})
have analyzed the Weinberg sum rules without the axial continuum.

In Fig. 1 we show our spectral density with axial continuum using the
same width as the vector continuum \cite{Shuryak_cor} and with a
threshhold value to be determined below. In this figure
the dashed curves correspond to the experimental uncertainty in the
branching ratio into $a_1$.  One can see that this is a rather
non-trivial, sign-changing function, which should obey the sum rules under
consideration.  Naturally, the new sum rule we consider is
more sensitive to the large $s$ behavior of the difference of the
spectral densities.  Thus we may at least ask whether {\it all} sum
rules are consistent with one common value of the parameter $E_A$.

In Fig. 2 (a-c) we have plotted sum rules I to III as functions of
$E_A$.  The horizontal dashed line shows in all cases
the right side of the sum rule which depends on the vacuum quark condensate
or $F_{\pi}$ as appropriate.  The intersection of the lines should
occur at the same value of $E_A$.  As explained in the previous section,
we do not know exactly the VEV of the relevent operators, therefore
we use the vacuum dominance estimate, with
\begin{equation}
|<0|\bar{u}u|0>|^{1/3} \, = \, 240 \, {\rm MeV} \,\,\,\,\,
(\mu=1 \,\,{\rm GeV}) \, .
\end{equation}
Fortunately, there seems to be
very little sensitivity to the value of the quark condensates.
One can clearly see that sum rules II and III are
quite consistent with the common value of $E_A$ = 1.45 GeV.
This observation is nontrivial.

Now we can come back to the first sum rule, use this value of $E_A$ as
input, and compare the numerical value of the integral to the right hand
side.  This procedure predicts $F_\pi$ about 5\% higher than the experimental
value.

Finally, let us comment on a closely related integral of the spectral
densities under consideration. It was shown in \cite{pionmass}
that the electromagnetic mass difference of pions can be expressed as
\begin{equation}
m^2_{\pi^+}-m^2_{\pi^0}={3 e^2 \over 16 \pi^2 F^2_\pi}
\int ds \, \ln \left({\Lambda^2+s \over s}\right)
[\rho_V(s)-\rho_A(s)] \, ,
\end{equation}
where $\Lambda$ is some cutoff parameter used to regulate the divergent
integral over virtual momentum in the loop. The result obviously depends
on it; only in one particular limit, namely, for $\Lambda \gg
m_\rho, m_{a_1}$ and for the original Weinberg values of the
$\rho$ and $a_1$ parameters without continuum one can get rid of it
and recover the original result $m^2_{\pi^+}-m^2_{\pi^0}=
(3\ln2\alpha/2\pi)m_{\rho}^2$ of \cite{pionmass}.
However, for the parameters extracted from data
as explained above, it is no longer true. The integral does
depend on the cutoff $\Lambda$.

In Fig. 2 (d) we show this sum rule with $\Lambda = 2$ GeV as a function of
$E_A$.  Note that the value of the
pion mass splitting is very sensitive
to $E_A$, and can even change sign if it is only 40 MeV above the
suggested value. However, at $E_A = 1.45$ GeV it agrees with the experimental
value (horizontal line) reasonably well.  Fine tuning could be accomplished
by adjusting the cutoff $\Lambda$, but we shall not do this.

\section{ Finite Temperature Sum Rules}

In this section we first generalize Weinberg's two sum rules to finite
temperature using essentially the same methods as he used without
any specific reference to QCD.  Then we verify the generalizations by
using the OPE, which also allows us to obtain the finite temperature
extension of sum rule III.  Finally, we investigate the behavior of
these sum rules at low temperature.

\subsection{Derivation of Weinberg-type sum rules at fixed momentum}

Consideration of Weinberg-type sum rules at finite temperature (or
chemical potential) is more involved than at zero temperature.
Lorentz invariance is not manifest because there
is a preferred frame of reference, the frame in which the matter is
at rest.  Thus spectral densities and other functions may depend on
energy and momentum separately and not just on their invariant $s$.
Also, the number of Lorentz tensors is greater because there is a new
vector available, namely, the vector $u_{\mu}$ = (1,0,0,0) which
specifies the rest frame of the matter.

For a given 4-momentum p it is useful to define two projection tensors.
The first one, $P_T^{\mu\nu}$, is both 3- and 4-dimensionally transverse,
\begin{equation}
P_T^{ij} \, \equiv \, \delta^{ij} \, - \, \frac{p^ip^j}{{\bf p}^2} \, ,
\end{equation}
with all other components zero.  The second one, $P_L^{\mu\nu}$,
is only 4-dimensionally transverse,
\begin{equation}
P_L^{\mu\nu} \, \equiv \, -\left( g^{\mu\nu} \, - \, \frac{p^{\mu}p^{\nu}}
{p^2} \, + \, P_T^{\mu\nu} \right) \, .
\end{equation}
The notation is $L$ for longitudinal and $T$ for transverse with respect
to ${\bf p}$. There are no other symmetric second rank tensors which
are 4-dimensionally transverse.

We now define the longitudinal and transverse spectral densities for
the vector current as
\begin{equation}
<V_a^{\mu}(x) V_b^{\nu}(0)> \, = \,
\frac{\delta^{ab}}{(2\pi)^3}
\int d^4p \, \theta(p^0) \, e^{ip\cdot x}
\left[ \rho_V^L P_L^{\mu\nu} \, + \, \rho_V^T P_T^{\mu\nu}
\right] \, ,
\end{equation}
and for the axial vector current as
\begin{equation}
<A_a^{\mu}(x) A_b^{\nu}(0)> \, = \,
\frac{\delta^{ab}}{(2\pi)^3}
\int d^4p \, \theta(p^0) \, e^{ip\cdot x}
\left[ \rho_A^L P_L^{\mu\nu} \, + \, \rho_A^T P_T^{\mu\nu}
\right] \, .
\end{equation}
In these expressions the angular brackets refer to the thermal average.
In general the spectral densities depend on $p^0$ and ${\bf p}$ separately
as well as on the temperature (and chemical potential).  These definitions
are standard and insure that both spectral densities are non-negative.
In the vacuum we can always go to the rest frame of a massive particle,
and in that frame there can be no difference between longitudinal and
transverse polarizations, so that $\rho_L = \rho_T = \rho$.  Since
$P_L^{\mu\nu}+P_T^{\mu\nu} = -(g^{\mu\nu} - p^{\mu}p^{\nu}/p^2)$
these equations collapse to eqs. (8-9).  The pion, being a massless
Goldstone boson, is special.  It contributes to the longitudinal axial
spectral density and not to the transverse one.  In fact, we could write
\begin{equation}
F_{\pi}^2 \delta(p^2) p^{\mu}p^{\nu} \, = \, F_{\pi}^2 p^2 \delta(p^2)
P_L^{\mu\nu} \, .
\end{equation}
This shouldn't be done at finite temperature because the contribution
of the pion to the longitudinal spectral density cannot be assumed
to be a delta function in $p^2$.  In general the pion's dispersion
relation will be more complicated and will develop a width at nonzero
momentum.  Therefore, we do not try to separate out the pionic
contribution but subsume it in the spectral density $\rho_A^L$,
without any loss of generality.

Following Weinberg, we define a three-point function by
\begin{equation}
-i\epsilon_{abc}M^{\mu\nu\lambda}(q,p) \, = \, \int d^4x\,d^4y
\, e^{-i(q\cdot x+p\cdot y)} <{\cal T}\left[ A_a^{\mu}(x)
A_b^{\nu}(y) V_c^{\lambda}(0) \right] > \, .
\end{equation}
We multiply both sides with $q_{\mu}$.  On the right side we can use
\begin{equation}
q_{\mu}e^{-i(q\cdot x+p\cdot y)} \, = \, i\frac{\partial}{\partial x^{\mu}}
e^{-i(q\cdot x+p\cdot y)} \, .
\end{equation}
Both the vector and axial-vector currents are conserved.  We assume
that we can integrate by parts and that the surface term is zero.
The nonzero contribution comes from
\begin{displaymath}
\frac{\partial}{\partial x^{\mu}} \left\{ {\cal T}\left[ A_a^{\mu}(x)
A_b^{\nu}(y) V_c^{\lambda}(0) \right] \right\}
\end{displaymath}
\begin{eqnarray}
&=& \delta(x^0-y^0)\left\{ \theta(x^0)\left[ A_a^0(x),A_b^{\nu}(y) \right]
V_c^{\lambda}(0) + \theta(-x^0)V_c^{\lambda}(0) \left[ A_a^0(x),
A_b^{\nu}(y) \right] \right\} \nonumber \\
&+& \delta(x^0)\left\{ \theta(y^0) A_b^{\nu}(y) \left[ A_a^0(x),
V_c^{\lambda}(0) \right] + \theta(-y^0) \left[ A_a^0(x),
V_c^{\lambda}(0) \right] A_b^{\nu}(y)\right\} \, .
\end{eqnarray}
{}From this expression we see the need for knowledge of the equal time
commutators.  Consistent with the normalization of eqs. (5-7) we have
\begin{eqnarray}
\delta(z^0)\left[A_a^0(x),A_b^{\nu}(y)\right] &=& i\epsilon_{abd}
V_d^{\nu}(x)\delta({\bf z})+S_{Vab}^{\nu j}({\bf x})
\frac{\partial}{\partial z^j} \delta({\bf z}) \, , \nonumber \\
\delta(z^0)\left[A_a^0(x),V_b^{\nu}(y)\right] &=& i\epsilon_{abd}
A_d^{\nu}(x)\delta({\bf z})+S_{Aab}^{\nu j}({\bf x})
\frac{\partial}{\partial z^j} \delta({\bf z}) \, .
\end{eqnarray}
Here $z = x - y$, and the $S$'s denote the Schwinger terms.

Consider now the contribution of the Schwinger terms to the thermal
average.  Generically they will be of the form
\begin{equation}
<SJ> \, = \, Z^{-1} \sum_{m,n}e^{-K_n/T} <n|S|m><m|J|n> \, ,
\end{equation}
where $K = H -\mu N$ is the Hamiltonian minus the chemical potential
times conserved particle number, the states are chosen to be eigenstates
of $H$, $N$, and isospin, and $J$ is either the vector or the axial-vector
current.  $J$ has isospin one, so we get zero if either (i) $S$ is a
c-number, or (ii) $S$ is an operator with no isospin one component.
We shall assume that one of these holds.  Then
\begin{displaymath}
\frac{\partial}{\partial x^{\mu}} < {\cal T}\left[ A_a^{\mu}(x)
A_b^{\nu}(y) V_c^{\lambda}(0) \right] > \, = \,
\end{displaymath}
\begin{equation}
i\epsilon_{abd}\delta(x-y)<{\cal T}\left[V_d^{\nu}(x)V_c^{\lambda}(0)
\right] > \, + \, i\epsilon_{acd}\delta(x)<{\cal T}
\left[A_b^{\nu}(y)A_d^{\lambda}(0) \right] > \, .
\end{equation}
It is now a simple matter to show that
\begin{equation}
\frac{1}{2}q_{\mu}M^{\mu\nu\lambda}(q,p) \, = \,
D_V^{\nu\lambda}(q+p) - D_A^{\nu\lambda}(p) \, ,
\end{equation}
where the $D$'s are the propagators for the currents, such as
\begin{equation}
\delta_{ab}D_A^{\nu\lambda}(p) \, = \, \int d^4y \, e^{-ip\cdot y}
<{\cal T} \left[ A_a^{\nu}(y)A_b^{\lambda}(0) \right] > \, .
\end{equation}
Similarly, one can show that
\begin{equation}
\frac{1}{2}(q+p)_{\lambda}M^{\mu\nu\lambda}(q,p) \, = \,
D_A^{\mu\nu}(q) - D_A^{\mu\nu}(p) \, .
\end{equation}
These Ward identities have exactly the same form as at zero temperature
\cite{Weinberg}.

With a similar consideration of the three-point function
\begin{equation}
-i\epsilon_{abc}N^{\mu\nu\lambda}(q,p) \, = \, \int d^4x\,d^4y
\, e^{-i(q\cdot x+p\cdot y)} <{\cal T}\left[ V_a^{\mu}(x)
V_b^{\nu}(y) V_c^{\lambda}(0) \right] > \, ,
\end{equation}
one can prove two more Ward identities,
\begin{equation}
\frac{1}{2}q_{\mu}N^{\mu\nu\lambda}(q,p) \, = \,
D_V^{\nu\lambda}(q+p) - D_V^{\nu\lambda}(p) \, ,
\end{equation}
and
\begin{equation}
\frac{1}{2}(q+p)_{\lambda}N^{\mu\nu\lambda}(q,p) \, = \,
D_V^{\mu\nu}(q) - D_V^{\mu\nu}(p) \, .
\end{equation}

Multiply eq. (42) by $(q+p)_{\lambda}$ and eq. (44) by $q_{\mu}$.  Do
the same for the other two Ward identities.  One obtains the constraints
\begin{equation}
(q+p)_{\lambda}D_V^{\nu\lambda}(q+p) \, = \,
q_{\lambda}D_V^{\nu\lambda}(q) \, + \,
p_{\lambda}D_V^{\nu\lambda}(p) \, = \,
q_{\lambda}D_A^{\nu\lambda}(q) \, + \,
p_{\lambda}D_A^{\nu\lambda}(p) \, .
\end{equation}
This implies linearity in the momentum,
\begin{equation}
k_{\lambda}D_V^{\nu\lambda}(k) \, = \,
k_{\lambda}D_A^{\nu\lambda}(k) \, = \,
C^{\nu\lambda}k_{\lambda} \, ,
\end{equation}
where $C^{\nu\lambda}$ is momentum-independent (but can depend on
temperature) and is the same for the vector
and axial-vector channels.  By taking the Fourier transform of these
relations we can find the thermal average of the equal time commutators,
\begin{equation}
\delta(x^0)<\left[ V_a^{\nu}(x),V_b^0(0) \right]> \, = \,
\delta(x^0)<\left[ A_a^{\nu}(x),A_b^0(0) \right]> \, = \,
\delta_{ab}C^{\nu\lambda}\frac{\partial}{\partial x^{\lambda}}
\delta(x) \, .
\end{equation}

The commutators above can be expressed in terms of the spectral
densities from eqs. (31-32).  Taking their {\it difference} one
obtains the finite temperature generalization of the first Weinberg
sum rule,
\begin{equation}
{\rm I} \,\,\,\,\,\,\,\,\,\,\,\,
\int_0^{\infty} \frac{d\omega \, \omega}{\omega^2-{\bf p}^2}
\left[ \rho_V^L(\omega,{\bf p})
-\rho_A^L(\omega,{\bf p}) \right] \, = \, 0 \, .
\end{equation}
Notice that this sum rule involves only the longitudinal spectral
densities and not the transverse ones.  At zero temperature
the spectral densities depend only on $p^2 = s = \omega^2 - {\bf p}^2$.
Then this equation reduces to eq. (1) once we remember to separate
out the pion piece of $\rho_A^L$, namely, $sF_{\pi}^2\delta(s)$.
At finite temperature, the spectral densities in general will depend
on $\omega$ and ${\bf p}$ separately and not just on the combination $s$.
Then this sum rule must be satisfied at {\it each} value of the
momentum.

At this point, Weinberg made an additional assumption in order to
obtain the second sum rule (eq. (2)): the currents behave like
free fields as $p^2 \rightarrow \infty$.  He also related the
difference between the vector and axial-vector propagators to the
matrix element of a particular operator between the vacuum and a
one pion state.  This is difficult to generalize to an ensemble
average.  To obtain the finite temperature generalization of the
second sum rule we follow the arguments of Das, Mathur and Okubo
\cite{Das1} instead.

Deleting the index $V$ or $A$ the explicit expressions for the
propagator and the Schwinger term are
\begin{eqnarray}
D^{00}(p^0,{\bf p})&=&{\bf p}^2D_L(p^0,{\bf p}) \, , \\
D^{0j}(p^0,{\bf p})&=&p^0p^jD_L(p^0,{\bf p}) \, , \\
D^{ij}(p^0,{\bf p})&=&\left(\delta^{ij}-\frac{p^ip^j}{{\bf p}^2}
\right)D_T(p^0,{\bf p})\,+\,\frac{p^ip^j}{{\bf p}^2}D_L'(p^0,{\bf p})\,,
\end{eqnarray}
where
\begin{eqnarray}
D_L(p^0,{\bf p})&=&2i\int_0^{\infty}\frac{d\omega\,\omega}{\omega^2
-{\bf p}^2} \left[ \frac{\rho^L(\omega,{\bf p})}
{\omega^2-p_0^2+i\epsilon} \right] \, , \\
D_L'(p^0,{\bf p})&=&2i\int_0^{\infty}\frac{d\omega\,\omega^3}{\omega^2
-{\bf p}^2} \left[ \frac{\rho^L(\omega,{\bf p})}
{\omega^2-p_0^2+i\epsilon} \right] \, ,  \\
D_T(p^0,{\bf p})&=&2i\int_0^{\infty}d\omega\,\omega \left[
\frac{\rho^T(\omega,{\bf p})}{\omega^2-p_0^2+i\epsilon} \right] \, ,
\end{eqnarray}
and
\begin{equation}
C^{00}\,=\,C^{0j}\,=\,C^{j0}\,=\,0 \, , \,\,\,\,\,\,\,\,\,
C^{ij}({\bf p})\,=\,\delta^{ij}D_S({\bf p}) \, ,
\end{equation}
where
\begin{equation}
D_S({\bf p})\,=\,2i\int_0^{\infty}\frac{d\omega\,\omega}{\omega^2
-{\bf p}^2} \, \rho^L(\omega,{\bf p}) \, .
\end{equation}

The first observation we can make concerns the thermally averaged Schwinger
term $C$.  Since it is the same for the vector and the axial-vector
correlators, by eq. (47), the $D_S({\bf p})$ must be the same as well.
Equating them reproduces the first finite temperature sum rule (eq. (49)).

The essence of the argument of Das, Mathur and Okubo is that spontaneous
chiral symmetry breaking is a low energy phenomenon.  At very high energy
it must disappear, at least in the limit that quark masses are zero
and chiral symmetry is exact.  Thus the difference between the vector
and axial-vector propagators should go to zero at very high energy,
\begin{equation}
\lim_{p^0 \rightarrow \infty, \,\,  {\bf p} \, {\rm fixed}} \,\,
\left[ D_V^{\mu\nu}(p^0,{\bf p}) \, - \,
D_A^{\mu\nu}(p^0,{\bf p}) \right] \,=\, 0 \, .
\end{equation}
If we do this for the time-time or time-space components of the
propagators, that is, for the $D_L$, we again reproduce the first
finite temperature sum rule.  Expanding to the next order in
$1/p_0^2$ we obtain a finite temperature generalization of the
second zero temperature sum rule, which is,
\begin{equation}
{\rm II-L} \,\,\,\,\,\,\,\,\,\,\,\,
\int_0^{\infty} d\omega \, \omega
\left[ \rho_V^L(\omega,{\bf p})
-\rho_A^L(\omega,{\bf p}) \right] \, = \, 0 \, .
\end{equation}
Like the first, this sum rule involves only the longitudinal spectral
densities, and we call it II-L.  Also like the first, it reduces
to the original Weinberg sum rule as the temperature and/or chemical
potential go to zero.

Next we consider the space-space components of the propagators.
Examination of the $D_L'$ in the infinite energy limit gives us
the sum rule II-L and nothing new.  Examination of the $D_T$
in the infinite energy limit gives us another sum rule which we
call II-T because it involves the transverse spectral densities,
\begin{equation}
{\rm II-T} \,\,\,\,\,\,\,\,\,\,\,\,
\int_0^{\infty} d\omega \, \omega
\left[ \rho_V^T(\omega,{\bf p})
-\rho_A^T(\omega,{\bf p}) \right] \, = \, 0 \, .
\end{equation}
The finite temperature sum rules II-L and II-T should become
degenerate at ${\bf p} = {\bf 0}$ because there ought not to be any
difference between longitudinal and transverse excitations at rest.
The sum rule II-T also then reduces to the original second sum
rule in the vacuum.

We want to emphasize that the sum rules derived in this section,
I, II-L and II-T, must be satisfied for {\it every} value
of the momentum.  Furthermore, our derivation is more general than
QCD; any theory which satisfies the assumptions we made must
obey these sum rules.  Perhaps they would be useful in the context
of models of the electroweak interactions where the Higgs particle
is a composite of other fields or for technicolor theories.

\subsection{Sum rules and the operator product expansion}

Application of the OPE to finite temperature has a peculiar history.
In the first papers (\cite{Bochkarev} and several later ones)
the authors considered only the
$T$-dependence of average values of the same
operators as at $T=0$, the Lorentz scalars. However, the rest frame
of the heat bath selects a 4-vector, thus symmetric tensors should also be
included.  In fact, the situation is completely
analogous to that in deep-inelastic scattering, for which one also has
a preferred frame, that of the target.  Thus, one can simply use formulae
derived in that context (see discussion in \cite{Shuryak_cor}).
The finite temperature sum rules were recently re-examined along these lines
in \cite{Hatsuda}.

The fact that we are not going to discuss vector and axial channels as such,
but only concentrate on their {\it difference}, brings in significant
simplifications. Most operators describing the interaction of a quark with
the gluonic field are chirality blind and therefore
cancel.  In the chiral limit, the difference appears only starting with
the four-quark operators.

To leading order in the momentum the difference between the vector and
axial-vector correlators is given by the OPE to be
\begin{eqnarray}
\Delta D^{\mu\nu}&=&-i\frac{4\pi\alpha_s}{(p^2)^3}
\mbox{\Large [} p^2<{\cal O}^{\mu\nu}>- \, p^\mu p_\alpha
<{\cal O}^{\alpha\nu}>
- \, p^\nu p_\alpha <{\cal O}^{\mu\alpha}> \nonumber \\
&+& g^{\mu\nu} p_\alpha p_\beta <{\cal O}^{\alpha\beta}> \mbox{\Large ]}
+{\rm order}\,(1/p^6) \, ,
\end{eqnarray}
where the operator ${\cal O}$ was defined in section 2.
This structure first appeared in the OPE analysis of the next-twist
correction to deep inelastic scattering in \cite{nexttwist}.
Observe that this quantity is transverse: $p_\mu \Delta D^{\mu\nu}=
p_\nu \Delta D^{\mu\nu}=0$.  This is consistent with eq. (47), the
equality of the Schwinger terms, and therefore with the assumptions
made to derive it.

First, consider $\Delta D^{00}$.  In terms of the spectral densities
it is given by eqs. (50) and (53).  Expand it in inverse powers of
$p_0^2$ in the limit that $|p_0| \rightarrow \infty$.  Since the
coefficients of $1/p_0^2$ and $1/p_0^4$ in eq. (61) are zero it
must be that the corresponding coefficients in eq. (50) are also
zero.  This gives us the finite temperature sum rules I and II-L
immediately.  We can say nothing about the next term without knowledge
of higher dimension operators in the OPE, which would contribute to order
$1/p_0^6$.

Next, consider $\Delta D^\mu_\mu$.  From eqs. (50-55) it is
\begin{equation}
\Delta D^\mu_\mu \,=\,2i\int_0^{\infty}\frac{d\omega\,\omega}
{p_0^2-\omega^2-i\epsilon} \left[ 2\Delta\rho^T(\omega,{\bf p})+
\Delta\rho^L(\omega,{\bf p}) \right] \, .
\end{equation}
Again, expand in inverse powers of $p_0^2$.  The term of order
$1/p_0^2$, when combined with the just derived sum rule II-L,
gives us the sum rule II-T.  The term of order $1/p_0^4$ gives us
the finite temperature version of sum rule III.
\begin{equation}
{\rm III} \;\;\;\;
\int_0^{\infty} d\omega\,\omega^3 \left[ 2\Delta\rho^T(\omega,{\bf p})+
\Delta\rho^L(\omega,{\bf p}) \right] \,=\,
-2\pi\alpha_s \left[ <{\cal O}^\mu_\mu> + 2<{\cal O}^{00}> \right] \, .
\end{equation}

We can make two observations about this sum rule.  In the limit of
vanishing temperature, Lorentz covariance says that
\begin{equation}
<{\cal O}^{00}>_{T=0} \,=\, \frac{1}{4} <{\cal O}^\mu_\mu>_{T=0} \, .
\end{equation}
This reduces eq.(63) to the previously derived zero temperature sum
rule eq. (16).  At finite temperature, the right side of eq. (63)
depends on $T$ but not on ${\bf p}$.  Therefore, the integral on
the left side must be momentum-independent.  If the integral
is known at zero momentum, for example, then it must have the same
value for any momentum.

\subsection{The low temperature limit}

As we are taking the zero quark mass limit in this work, the pion is massless
below any critical temperature for chiral symmetry restoration and/or
deconfinement, and thus at
parametrically low temperature the heat bath is dominated
by pions.
In \cite{Eletsky_etal} the so-called
Dey-Eletsky-Ioffe mixing theorem was proven, which says that,
to order $T^2$, there is no change in the masses of vector and axial-vector
mesons.  What changes are the couplings to the currents.  The finite
temperature correlators can be described by a mixing
between the vector and axial-vector $T=0$ correlators with a temperature
dependent coefficient,
\begin{eqnarray}
D_V^{\mu\nu}(p,T)&=&(1-\epsilon)D_V^{\mu\nu}(p,0)\,+\,
\epsilon D_A^{\mu\nu}(p,0) \, , \\
D_A^{\mu\nu}(p,T)&=&(1-\epsilon)D_A^{\mu\nu}(p,0)\,+\,
\epsilon D_V^{\mu\nu}(p,0) \, .
\end{eqnarray}
These are valid to first order in $\epsilon \equiv T^2/6F_{\pi}^2$.
This implies the same mixing of the spectral densities, namely,
\begin{eqnarray}
\rho_V(p^0,{\bf p},T)&=&(1-\epsilon)\rho_V(s,0)\,+\,\epsilon
\rho_A(s,0) \, , \\
\rho_A(p^0,{\bf p},T)&=&(1-\epsilon)\rho_A(s,0)\,+\,\epsilon
\rho_V(s,0) \, ,
\end{eqnarray}
with the appropriate longitudinal and transverse subscripts.
The temperature dependence of the pion decay coupling was thus
proven to be $F_{\pi}^2(T) = (1-\epsilon)F_{\pi}^2$ for small $T$
consistent with the prediction of chiral perturbation theory \cite{Gasser}.
Therefore, the finite temperature sum rules I (eq. (49)),
II-L (eq. (59)) and II-T (eq. (60)) reduce to the original,
zero temperature sum rules but with both sides of the eqs. (1) and
(2) multiplied by the factor $1-2\epsilon$.

One may ask whether the third sum rule also obeys the Dey-Eletsky-Ioffe
mixing theorem.  A general formula describing the thermal average of any
four-quark operator using soft pion methods was derived in \cite{Eletsky}.
For an operator ${\cal O}_{AB} = \bar q A q \bar q B q$ the expression is
\begin{eqnarray}
<{\cal O}_{AB}>&=&{<\bar u u >^2 \over 144}\left(1-{T^2\over 4 F_\pi^2}
\right)\left[{\rm Tr}(A) {\rm Tr}(B) - {\rm Tr}(AB)\right]
\nonumber \\ &-&
{<\bar u u >^2 \over 144}{T^2\over 12 F_\pi^2}
\left[{\rm Tr} (\gamma_5 \tau^a A)
{\rm Tr}(\gamma_5 \tau^a B) - {\rm Tr}
(\gamma_5 \tau^a A \gamma_5 \tau^a B)\right] \, ,
\end{eqnarray}
where it is assumed that at $T=0$ one can use the vacuum dominance
approximation.

The average of the four-quark operator appearing in sum rule III gets
multiplied by the correct factor $1-2\epsilon$, as shown by Eletsky
\cite{Eletsky}.  This is not a trivial result:
the average value of an arbitrarily chosen four-quark operator will
not have the same temperature dependence.
As already emphasized by Eletsky,
a simplistic application of factorization at nonzero temperature,
which would suggest the same behavior as for the quark-condensate squared,
\begin{equation}
<\bar u u>^2 \,=\, \left(1-{T^2 \over 4 F^2_\pi}\right)<0|\bar{u}u|0>^2,
\end{equation}
would be wrong, and in fact violates the sum rule.

In summary, at low temperature the sum rules under discussion satisfy the
Dey-Eletsky-Ioffe mixing theorem exactly.

\section{Scenarios for Chiral Symmetry Restoration}

Chiral transformations are rotations of the quark field with
$\gamma_5$, and they
may or may not have the  $SU(N_f)$ (isospin) generators.
The corresponding $U(1)_A$ and $SU(N_f)_A$ have different fates in QCD;
the former is explicitly violated by the anomaly, the latter is broken
spontaneously at low temperature and is restored at some critical
temperature $T_c$, provided the quark mass is
stricly zero as it is assumed in this paper.
However, as the $\rho$ and $a_1$ channels are both isospin-1,
the symmetry which can mix them is $U(1)_A$.

If both chiral symmetries are restored, one may
conclude that left and right-polarized quarks form completely
independent subsystems. If so, quarks never change chirality, and
there should be no difference between vector and axial-vector correlators,
and the object of our consideration is simply zero.

This is indeed expected to happen at very high temperatures,
but in the critical region $T\approx T_c$ all depends on
the mechanism of the $U(1)_A$ symmetry breaking.
One of us recently wrote
a mini-review on this subject \cite{Shuryak_U(1)}, and it is probably enough
to mention here that it is most likely that
the $U(1)_A$ symmetry is practically restored at $T\approx T_c$.
More specifically, there is direct
evidence from lattice simulations that the difference between
vector and axial-vector correlators
do indeed vanish around this point, to within the accuracy of the calculation.

In this section we speculute on exactly how
this difference goes to zero with increasing temperature.
Generally, one may suggest many different scenarios.  Let us discuss
the following three.

\subsection{Mixing of vector and axial-vector spectral densities}

The simplest scenario is that the $T$-dependence factorizes.
It means that the vector and axial-vector spectral
densities  mix, without changing their shape, as in the
low temperature limit considered in the previous section,
only with a more general function $\epsilon(T)$. When
the mixing becomes maximal, $\epsilon = 1/2$, chiral symmetry is
restored.

It is amusing to see at what temperature this occurs using the lowest order
formula, $\epsilon=T^2/6F_\pi^2$.
This estimate gives $T_{\rm complete \; mixing} = \sqrt{3}F_{\pi} \approx 164$
MeV, which is indeed roughly equal to the expected critical temperature $T_c$.

\subsection{Shift in meson pole masses and residues}

In this scenario we assume that the $\rho$ and $a_1$ mesons
retain their identities and dominate the correlation function.
However, their parameters change with temperature. In particular,
the masses may move towards each other \cite{Song},
or go to zero \cite{Brown_Rho}.
At $T_c$ they become degenerate, and chiral symmetry is restored.

It is instructive then to look at the sum rules.
Let us assume that vector meson dominance is a good approximation
for the spectral densities and not worry about the continuum contribution
for the time being.  Let us focus on zero momentum for the sake of
simplicity.  When a pole mass is defined at finite temperature, it
is usually defined as the energy of the excitation at zero momentum.

The vector spectral density is (there is no difference between
longitudinal and transverse at zero momentum)
\begin{equation}
\rho_V(\omega) \,=\, \frac{1}{\pi}\,\frac{m_{\rho}^4}{g_{\rho}^2}
\, {\rm Im} \,
\frac{1}{\omega^2-m_{\rho}^2-\Pi_R^{\rho}(\omega)-
i\Pi_I^{\rho}(\omega)} \, ,
\end{equation}
where $\Pi_R^{\rho}$ and $\Pi_I^{\rho}$ are the real and imaginary
parts of the $\rho$ self-energy at temperature $T$.  In the narrow
width approximation this becomes
\begin{equation}
\rho_V(\omega) \,=\, \frac{m_{\rho}^4}{g_{\rho}^2} \delta \left(
\omega^2-m_{\rho}^2-\Pi_R^{\rho}(\omega) \right) \, .
\end{equation}
The pole mass is determined self-consistently from $m_{\rho}^2(T)
= m_{\rho}^2 + \Pi_R^{\rho}(m_{\rho}(T))$.  Then the spectral density
can be rewritten as
\begin{equation}
\rho_V(\omega) \,=\, Z_{\rho}(T) \frac{m_{\rho}^4}{g_{\rho}^2}
\delta \left( \omega^2-m_{\rho}^2(T) \right) \, ,
\end{equation}
where the temperature-dependent residue is
\begin{equation}
Z_{\rho}^{-1}(T) \,=\, \left| 1 - \frac{d}{d\omega^2}\Pi_R^{\rho}(\omega)
\right| \, .
\end{equation}
The normalization is $Z_{\rho}(0) = 1$.  Similarly
\begin{equation}
\rho_A(\omega) \,=\, Z_{a}(T) \frac{m_{a_1}^4}{g_a^2}
\delta \left( \omega^2-m_{a_1}^2(T) \right) \,+\,
Z_{\pi}(T) F_{\pi}^2 \omega^2
\delta \left( \omega^2 \right) \, .
\end{equation}

Substituting these spectral densities into the finite temperature
sum rules I and II-L/II-T tells us that the $\rho$ and $a_1$
residues are equal
\begin{equation}
Z_{\rho}(T) \,=\, Z_a(T) \, ,
\end{equation}
and that the pion residue is
\begin{equation}
Z_{\pi}(T) \,=\, 2Z_{\rho}(T) \left[ \frac{m_{\rho}^2}{m_{\rho}^2(T)}
-\frac{m_{\rho}^2}{m_{a_1}^2(T)} \right] \, .
\end{equation}
We expect that $m_{a_1}^2(T) - m_{\rho}^2(T) \rightarrow 0$ as
the temperature increases.  Three types of behavior can be distinquished:
both the $\rho$ and the $a_1$ masses decrease with $T$, both masses
increase with $T$, or the $\rho$ mass increases while the $a_1$ mass
decreases with $T$.  The sum rules do not appear to rule out any of
these possibilities.  In any case, the result is that $Z_{\pi}(T)
\rightarrow 0$ unless $Z_{\rho}(T) \rightarrow \infty$, which seems
rather unphysical.

\subsection{Resonance broadening and downward shift of the continuum}

As distinct from the previous scenarios, it may be that particles
are not well-defined as we approach a chiral symmetry restoring
phase transition.  That is, the imaginary part of the self-energy
may become larger with increasing temperature.  This
broadening would also decrease the maximum peak value of the spectral
density.  Euphemistically, the vector and axial-vector mesons melt away.
There may be also a decrease in the threshholds $E_V(T)$
and $E_A(T)$ of the continuum.  The continuum would merge with the
broadened particle poles to give a very broad distribution of strength
in the spectral densities.  The difference of spectral densities
shown for $T = 0$ in Fig. 1 would become flatter and decrease everywhere
towards zero, effectively restoring chiral symmetry.

Concluding this section, we say once more that the sum rules by themselves
cannot of course tell which scenario is preferable.  However, the sum rules
can be used to significantly restrict the parametrization of the spectral
densities at nonzero temperature.

\section{Conclusion}

In this paper we studied Weinberg-type sum rules at zero
and at nonzero temperature.  All considerations were made in the
exact chiral limit of QCD, $m_q\rightarrow 0$.
In the former case, we derived a new sum rule of the Weinberg-type.
Although it belongs to an infinite series of sum rules,
one for each type of OPE term at small distances,
we think it is special in several respects. First, it is
relatively simple theoretically
because it is related to the VEV of a specific four-quark operator.
Sum rules of higher order than the third are much more complicated.
Second, it is related to the leading nonzero $\ln(\tau)$ term
of the correlators, while others can be related to sub-leading
terms which are much more difficult to single out, especially in
lattice simulations.

Continuing the zero temperature analysis, we re-examined the experimental
data together with all relevant sum rules.
We found that, although we do not have sufficient
information on the VEV of the operator for sum rule III,
we still can use it, together with sum rule II,
to fix the numerical value of $E_A$,
the continuum threshold in the axial channel. This
essentially closes the gap in the
experimental data, and allows one to test the original
Weinberg sum rules without any {\it ad hoc} assumptions.
Good agreement with the experimental values of $F_\pi$ and
the electromagnetic mass difference of pions provides a
non-trivial consistency check of the data used.

Our finite temperature analysis consists of several different parts.
First, following Weinberg's original derivation, one can find
generalizations of his sum rules to nonzero temperature.  Sum rule I
involves only the difference of the longitudinal spectral densities,
while sum rule II bifurcates into two sum rules, one involving the
longitudinal spectral densities, and the other involving the transverse
ones.  {\it These sum rules must be satisfied at each value of the momentum}.
These new features arise because of the appearance of a preferred
reference frame at nonzero temperature.  These sum rules were derived
without specific reference to QCD so they are applicable to other
theories satisfying the assumptions made.  We were also able to derive
them from the OPE.  Furthermore, we used the OPE to obtain the
finite temperature generalization of the new sum rule III, which makes
specific reference to the dynamics of QCD.

We also considered very low temperatures at which chiral
perturbation methods predict the general behavior of the correlators.
We showed that these results are in exact agreement with all sum rules under
consideration.

We would like to emphasize that the average value of the four-quark
operator which appears in sum rule III
is of great theoretical interest. It shows correlation between
densities and currents made of left and right-handed quarks and
is, in a sense, an order parameter for restoration of $U(1)_A$
chiral symmetry.  The average value in the QCD vacuum and
and at finite temperature can and should be studied in lattice numerical
simulations.  This task is facilitated by the fact that it does not
have any perturbative contributions.

Finally, we speculated on possible scenarios of chiral symmetry restoration.
We have no preferences among them, and only future work, including especially
lattice numerical simulations, can clarify which of them (if any) is
realized in QCD. However,
the derived sum rules should hold in any case,
thus providing some relations among parameters of the vector
and axial-vector spectral densities.

\section*{Acknowledgments}

J. I. K. was motivated in part by a seminar by Chungsik Song and the
ensuing questions by L. D. McLerran.
This work was supported by the U.S. Department of Energy under grants
DOE/DE-FG02-87ER40328, DOE/DE-FG02-88ER40388, and DOE/DE-FG02-93ER40768,
and by the U.S. National
Science Foundation under grant PHY89-04035.
The work was done during the program {\it Strong Interactions at Finite
Temperature}; we are much indebted to the ITP at Santa Barbara for
hospitality and support.

\newpage
\section*{Figure Captions}

\noindent Fig. 1.  The difference between the spectral densities of the
vector and axial-vector currents versus $s$ at zero temperature.
The two dashed curves show the uncertainty due to the
experimental determination of the $a_1$ coupling constant, described in the
text.  The abrupt change at $s$ = 0.8 GeV$^2$ corresponds to the sharp
onset of the $a_1$ contribution at $(m_\rho+m_\pi)^2$.\\

\noindent Fig. 2.  Dependence of the zero temperature sum rules I-III
on the effective perturbative threshhold $E_A$ in the axial channel.
The last panel shows the $\pi^+ - \pi^0$ mass difference.
As in Fig. 1, the solid and the
two dashed curves correspond to the central value and uncertainty in the $a_1$
coupling constant.
In all cases, the expected magnitude of the corresponding sum is shown by
the horizontal long-dashed line.

\end{document}